%% file: bare_jrnl.tex
\documentclass[lettersize,journal]{IEEEtran}
\usepackage{amsmath,amsfonts}
\usepackage{algorithm}
\usepackage{array}

\usepackage{cite}

\usepackage{amssymb}
\usepackage{amstext,mathrsfs}

\usepackage{algorithmicx} 
\usepackage{algpseudocode}

\usepackage{graphicx}
\usepackage{textcomp}
\usepackage{xcolor}
\def\BibTeX{{\rm B\kern-.05em{\sc i\kern-.025em b}\kern-.08em
		T\kern-.1667em\lower.7ex\hbox{E}\kern-.125emX}}

\usepackage{comment}

\begin{document}

\title{A Deep Reinforcement Learning based Scheduler for IoT Devices in Co-existence with 5G-NR}

\author{Shahida Jabeen

\thanks{This research was supported by the Natural Sciences and Engineering Research Council of Canada (NSERC). 
\textit{Corresponding author:} shahida.jabeen@uwaterloo.ca}

}

\maketitle

\begin{abstract}
Co-existence of 5G New Radio (5G-NR) with IoT devices is considered as a promising technique to enhance the spectral usage and efficiency of future cellular networks. In this paper, a unified framework has been proposed for allocating in-band resource blocks (RBs), i.e., within a multi-cell network, to 5G-NR users in co-existence with NB-IoT and LTE-M devices. First, a benchmark (upper-bound) scheduler has been designed for joint sub-carrier (SC) and modulation and coding scheme (MCS) allocation that maximizes instantaneous throughput and fairness among users/devices, while considering synchronous RB allocation in the neighboring cells. A series of numerical simulations with realistic ICI in an urban scenario have been used to compute benchmark upper-bound solutions for characterizing performance in terms of throughput, fairness, and delay.
Next, an edge learning based multi-agent deep reinforcement learning (DRL) framework has been developed for different DRL algorithms, specifically, a policy-based gradient network (PGN), a deep Q-learning based network (DQN), and an actor-critic based deep deterministic policy gradient network (DDPGN). The proposed DRL framework depends on interference allocation, where the actions are based on inter-cell-interference (ICI) instead of power, which can bypass the need for raw data sharing and/or inter-agent communication. The numerical results reveal that the interference allocation based DRL schedulers can significantly outperform their counterparts, where the actions are based on power allocation. Further, the performance of the proposed policy-based edge learning algorithms is close to the centralized ones.
\end{abstract}


\input{Introduction}

\input{Framework}
\input{Deep-Learning}

\input{Deep-Learning-Algorithms}
\input{Results}

\section{Conclusion and Future Work}
\label{sec:conclusion}
Design of an efficient, fair, and ICI-aware resource allocation scheme is essential for providing satisfactory QoS to ICI limited dense cellular IoT networks. In this paper, a unified, fair, and ICI-aware benchmark scheduler has been developed with joint SC and MCS allocation for NB-IoT, LTE-M, and 5G-NR users/devices deployed within a 5G-NR carrier. The proposed benchmark (or upper-bound) solutions can significantly outperform the fixed power budget based baseline scheduling solutions. However, due to significant computational requirements (especially with varying channel conditions) they can only be used by ISPs for benchmarking and planning 5G NR networks with co-existing and drastically growing IoT traffic.  

A unified DRL framework has been developed for training and testing different centralized as well as edge learning based DRL models for efficiently scheduling cellular IoT devices in co-existence with 5G-NR users. In particular, three DRL algorithms (i.e., {\bf \emph{DQN-IA, PGN-IA, DDPGN-IA}}) have been proposed, where the actions are designed to allocate interference instead of power. Numerical results show that they perform better than their power allocation based counterparts (i.e., {\bf \emph{DQN-PA, PGN-PA, DDPGN-PA}}). Specifically, the policy-based DRL algorithms can significantly outperform value-based DQN algorithms in terms of average sum throughput, fairness (via geometric mean throughput), and delay (via harmonic-mean throughput). Further, they can provide better adaptability against varying channel conditions. 

\bibliography{bare}

\end{document}

%% file: Introduction.tex
\section{Introduction}
\label{sec:intro} 
\IEEEPARstart{C}{ellular} IoT networks have been adopted widely across the globe to support the ever-increasing number of connections (i.e., in the licensed band) enabled by narrowband IoT (NB-IoT) and LTE for machines (LTE-M) standards \cite{NB-IoT-Primer, MTC-5G}. Although these 3GPP standardized technologies have complementary capabilities, they are envisioned to offer better performance than other low power wide area (LPWA) networks in terms of coverage, quality of service (QoS), scalability, and flexibility. It is predicted that in the next few years, a massive number of NB-IoT and LTE-M devices will be connected to cellular networks to address a wide spectrum of use cases (such as, broadband or mission critical applications). According to \cite{Ericsson2023}, by 2028 there will be 6 billion wide-area IoT connections (with 5.4 billion cellular IoT connections), given that the broadband and critical IoT reached 1.5 billion connections in 2022. 

Due to long-term deployment of typical IoT devices, they need to co-exist with other 5G and beyond radio access technologies for more deployment flexibility and higher energy efficiency over LTE \cite{GSMA-5G-Mobile-IoT}. Although, the applications for NB-IoT and LTE-M devices differ from each other, both of them are low-power radio access technologies requiring low bandwidth and thus can greatly benefit from spectral efficiency enhancements via 5G-NR co-existence. However, in order to schedule NB-IoT or LTE-M devices within a 5G-NR carrier, a number of design challenges need to be considered, such as, coverage, energy consumption, latency, and interference; peculiarly in urban cellular networks, where IoT carriers are deployed within a 5G-NR carrier, the allocation of resource blocks (RBs) using a sub-carrier (SC) granularity leads to many scenarios where the 5G transmissions wholly or partially overlap with the ones in neighboring cells, resulting in severe inter-cell interference (ICI) \cite{RRM-NB-IoT, RRM-NB-IoT-FD}. To mitigate such scenarios, the IoT carriers can be deployed in either an in-band or a guard-band mode. Nonetheless, the in-band operation also leads to significant ICI (on both downlink and uplink) due to the neighboring cells' IoT devices (in the case of synchronous RB allocation) or from 5G-NR UEs (in the case of asynchronous RB allocation); this is particularly true for urban and dense IoT networks. 

The ICI problem within an urban cellular network \cite{Deep-Learning-for-RM-in-IoT} requires sophisticated power allocation and interference management for IoT devices. Since, they are heterogeneous with respect to power, delay, and throughput limitations. The potentially varying physical channel and network characteristics require dynamic power and resource allocation schemes with ICI coordination for energy efficiency, throughput enhancement, latency reduction, coverage extension, as well as ICI mitigation in real-time. Note that the impact of ICI on the uplink is more severe as the instantaneous throughput directly affects the transmission delay and the number of re-transmissions; more delay for IoT devices result in more energy consumption and waiting time before going into their sleep mode.  

Typically, the rate adaptation window for ICI compensation is quite slim for sensors and actuators, where a small delay in data transmission could lead to data inconsistencies and signaling delays, respectively. To address these challenges, machine learning (ML) can play a pivotal role in real time power and resource allocation for large-scale IoT networks. The abundance of data and significant improvements in capabilities of modern hardware have made ML algorithms more realizable in real-time; where, individual state-of-the art RRM procedures can be replaced by autonomous ML algorithms. 

\subsection{Traditional Schedulers for IoT Devices}
The traditional schedulers for cellular or IoT networks like \cite{5G-RRM-for-IoT,NB-IoT-Scheduler,NB-IoT-Scheduler-Journal-paper,NB-IoT-Uplink-Scheduling-1,NB-IoT-Scheduling}, do not consider the impact of ICI rather the main focus has been on satisfying QoS requirements, improving resource utilization, and reducing energy consumption for IoT devices within a single cell. For instance, a local hybrid uplink scheduler has been designed in \cite{5G-RRM-for-IoT} for a next generation node (gNodeB) with multiple IoT devices in co-existence with LTE users. The proposed RB allocation scheme maximizes the bandwidth utilization without considering ICI. A downlink heuristic based scheduler, within a single cell environment, has been investigated in \cite{NB-IoT-Scheduler, NB-IoT-Scheduler-Journal-paper} to minimize the number of required resources (timeslots), while satisfying each device’s data requirement. Another single cell scheduler has been designed in \cite{NB-IoT-Uplink-Scheduling-1} to guarantee QoS requirements, while minimizing energy consumption for the IoT devices. A resource mapping scheme with fair allocations has been proposed in \cite{NB-IoT-Scheduling} that can achieve various trade-offs between power consumption, search space utilization, and fairness, while neglecting ICI. 

Some multi-cell IoT schedulers \cite{User-Scheduling-and-BS-Offloading} assume that each device is transmitting using a fixed power level and interference can be measured exactly via uplink training. In contrast, \cite{RRM-for-Cellular-IoT} investigates the presence of inter-band interference due to spectral leakage, where a joint non-convex NP-hard RB and power allocation problem has been formulated and sub-optimally solved to maximize the sum of device throughputs. A similar study in \cite{RRM-NB-IoT, RRM-NB-IoT-FD} highlights RB allocation issues that need to be resolved for achieving maximum spectral efficiency for NB-IoT devices; where, an intractable joint RB allocation framework has been formulated to mitigate the impact of ICI, reduce re-transmissions, and accommodate asymmetric traffic. Here, the joint RB allocation problem was solved sub-optimally using an iterative algorithm.

\subsection{ML based Schedulers for IoT Devices}
The traditional schedulers for cellular or IoT networks require mathematical models that must be analytically and computationally tractable to study their performance and effectiveness. Therefore, ML based tools/techniques have become popular recently to achieve near-optimal performance with affordable computational complexity. However, development of ML models for IoT networks is still challenging, mainly because there is no single model that could suffice the QoS requirements of a diverse range of IoT devices. Further, the reliability of the underlying ML model is essential for some IoT applications that are vulnerable to delay and throughput. 

In a nutshell, ML models must be efficiently crafted and unified to meet the QoS requirements of the underlying radio access technology and the characteristics unique to it. To this end, the existing ML models are generally categorized into two groups: supervised learning and DRL. 

In supervised learning, the model is trained with an input-output labeled dataset, thus, it can be used as universal function approximator for an already known function. For instance, supervised learning has been applied in \cite{ayala2019vrain}, where a trained deep neural network (DNN) is used to accelerate processing speed and reduce performance loss. To further improve the performance of DNNs, an ensemble of DNNs has been proposed in \cite{foukas2019iris}. The main advantage of using DNN for approximating a function is that, once trained, a DNN has a very low real-time computational cost, whereas using the original function might incur a much higher cost. However, the supervised learning approach is not feasible for IoT due to the complexity of universal function approximation and also the need for very large training datasets. 

In DRL based models, the desired output is not known beforehand, and an agent interacts with an environment and receives a reward/feedback, where the goal is to maximize the numerical value of the reward. DRL has two phases, the first one is an offline DNN construction phase, which correlates the value function with corresponding states and actions in the environment. The second one is an online deep Q-learning (DQL) phase, responsible for action selection and network updation. DRL seems more appropriate for scheduling IoT as it neither requires huge amount of labeled data nor a function that can label the data, but a performance metric that can be used to measure the quality of decision made by the agent, such as, a throughput or delay metric. DRL is particularly suitable for solving problems with very large input spaces, because a DNN algorithm learned properly through DRL can perform well for input instances that have never been tried in the learning phase (also known as the generalization capability of neural networks). 

The DRL based models can learn a wide range of complex functions using suitable neural network structures from a relatively small subset of input instances. For example, by adding more fully connected layers to a neural network, one can generally enhance its capability of representing more complex functions. With more layers, a neural network will have longer forward propagation time, thus, while addressing problems with real-time requirements, there will be a design trade-off between the functional representation capability of a neural network and its computational time cost.

\subsection{Main Contributions} 
The existing DRL based schedulers \cite{nasir2020deep, ghadimi2017reinforcement, awan2018robust,zafaruddin2019distributed, chinchali2018cellular, xu2017deep, sun2017learning, ahmed2019deep, xiong2018parametrized} mainly focus on value-based algorithms \cite{amiri2018machine, simsek2014learning} to solve either sum-rate maximization or proportionally fair scheduling problems for a very small network size \cite{khan2020centralized}. It is fair to say that the literature on the DRL based schedulers for IoT networks is limited and ICI-limited urban cellular 5G-NR networks with IoT co-existence have not been investigated before. 

In this paper, a unified framework has been developed for uplink\footnote{Note that a similar framework can be developed for the downlink.} scheduling within an ICI limited multi-cell network, while aggregating different radio access technologies, such as, NB-IoT, LTE-M, and 5G-NR. The proposed framework exploits a benchmark (upper-bound) scheduler for solving the joint SC and MCS allocation problem that maximizes instantaneous uplink throughput and fairness among IoT devices and 5G-NR users. A unified, fair, and ICI-aware scheduler for IoT devices deployed in-band with 5G-NR users has not been studied before. 

Numerical simulations in \cite{WF-5G-2021}, with realistic ICI parameters, show that the solutions of the benchmark (upper-bound) scheduler are close to the optimal and the fixed power budget based schedulers are far-off from the optimal.

Next, a multi-agent DRL based framework has been developed for finding feasible solutions to the benchmark (upper-bound) problem in an efficient manner, specifically the policy-based DRL algorithms like deep deterministic policy gradients (DDPG) have been exploited in the context of IoT networks, which are based on the notion of “actor” and “critic”; where the actor is a deep function approximator that takes an observed input instance and outputs an action and the critic is another deep function approximator that predicts the expected return of an action under a given input instance. 

In this paper, three DRL algorithms have been designed for solving the joint SC and MCS allocation problem, i.e., a deep Q-learning network ({\bf \emph{DQN-IA}}), a policy gradient network ({\bf \emph{PGN-IA}}), and a deep deterministic policy gradient network ({\bf \emph{DDPGN-IA}}); where, instead of taking actions based on power, the actions are taken based on ICI, which (unlike power) can be discretized easily. The results are obtained via realistic numerical simulations over both centralized learning as well as edge learning network architectures, which reveal that centralized {\bf \emph{DDPGN-IA}} can significantly outperform the fixed power budget based schedulers as well as other DRL algorithms in terms of sum-rate as well as proportionally fair throughput and delay. Further, the performance of the proposed policy-based edge learning algorithms, that do not share raw data and communicate with other agents, is close to the centralized algorithms that require inter-agent communication.

The remainder of this paper is organized as follows: Section~\ref{sec:framework} describes the 5G-NR framework and the joint SC and MCS allocation problem. Section~\ref{sec:DRL} has been dedicated to the proposed DRL framework, where the design of the existing DRL models to solve the joint SC and MCS allocation problem has been discussed. The network simulation setup is explained in Section~\ref{sec:setup}, while numerical results for the centralized as well as the edge learning based DRL framework have been compared and analyzed in Section~\ref{sec:results}. The final conclusions and future insights are presented in Section~\ref{sec:conclusion}.

%% file: Framework.tex
\section{5G-NR Framework}
\label{sec:framework}
The proposed 5G-NR framework is based on a co-existing multi-cell network of IoT devices and 5G-NR users. The IoT devices in the network communicate with their serving gNodeBs using a dedicated RB placed in-band within a 5G-NR carrier. Since, in-band IoT sub-carriers (SCs) \cite{NB-IoT-Interference} suffer from an increased co-channel interference even in synchronous multi-cell networks\footnote{Note that even with RB blanking, co-channel interference due to 5G-NR control symbols can degrade IoT link performance.}, the NB-IoT/LTE-M sub-carrier grids are assumed to be aligned with each other in order to ensure orthogonality and eliminate any inter-sub-carrier interference. 

Further, the dedicated RB allocation for IoT devices is assumed to be synchronized within all cells in the network. Although, the flexibility in 5G-NR significantly facilitates scheduling resources in time domain as well as frequency domain with different SC bandwidth and spacing, for simplicity, equal SC bandwidth and spacing for all RBs has been used within the 5G-NR carrier to achieve the best co-existence performance \cite{LTE-M-NB-IoT-Coexistence-5G-NR,NB-IoT-5G-Performance} with NB-IoT and LTE-M devices. 

The proposed framework consists of a set of fixed gNodeBs (i.e. one per cell) denoted by $\mathscr{B}$, where each one is equipped with a single antenna and is assumed to have unlimited back-haul capacity. The set of IoT devices are distributed uniformly over each cell with a single antenna and fixed transmit power budget ($P_{IoT}$). It is implicitly assumed that all 5G-NR gNodeBs are capable of periodically transmitting primary synchronization signal (PSS) and secondary synchronization signal (SSS) signals as well as narrowband PSS (NPSS) and narrowband SSS (NSSS) for LTE-M and NB-IoT devices, respectively; these signals are required to perform cell search including time/frequency synchronization and cell identity detection. 
\vspace{-0.2cm}
\subsection{Link Rate Model}
\label{sec:rate-model}
A \emph{realization}-based approach \cite{snapshot} has been used for computing link rates across multiple cells over each synchronized RB for both IoT devices as well as the 5G-NR users, where each realization $\omega \in \Omega$ of the network corresponds to a set of SCs ($\mathscr{S}$) and a set of time-slots ($t \in \mathscr{T}$). The set of users/devices connected to each gNodeB ($b\in \mathscr{B}$) is denoted by $I_b(\omega)$, and the user/device and gNodeB pair is denoted by $(i,b)$. All users/devices are assumed to be active (i.e., have uplink messages ready to be sent) and are connected to the same gNodeB for the duration of each realization.

During time slot $t$, the set of independent SC channel gains between the gNodeB $b$ and the user/device $i$ on the uplink are denoted by $\{G_{(i,b),b}(\omega,t)\}_{(i,b) \in (I_b(\omega),\mathscr{B})}$. The SC channel gains are assumed to be known beforehand and are flat over each RB for a given gNodeB and user/device pair with only small-scale fading parameters being changed for each timeslot; since, IoT device locations are fixed, the gains do not change drastically during each realization. A SC in each cell is allocated to at most one user/device and a device/user cannot have multiple uplink SCs assigned to it. Therefore, for the duration of each realization, it is reasonable to assume that the SC allocations will remain static. 

The uplink signal-to-interference-plus-noise ratio (SINR) from device $(i,b)$ to gNodeB $b$ on SC $s$ in timeslot $t$ of realization $\omega$ is define as follows: 
\begin{equation}
\label{eq:ULsinr}
\begin{aligned}
\gamma_{(i,b)}^{s}(\omega,t)&:= \frac{ {P}_{(i,b)}^{s} \times {G}_{(i,b),b}(\omega,t) }{N_{0} + \sum\limits_{(i',b') \in (I_b(\omega),\mathscr{B})}^{(i',b') \neq (i,b)} P_{(i',b')}^{s}  G_{(i',b'),b}(\omega,t))}
\end{aligned}
\end{equation}
Here, $P_{(i,b)}^{s}$ and $\boldsymbol{N_0}$ are the transmit power per SC and the average noise power per SC, respectively. 

The proposed framework uses a link adaptation process \cite{NB-IoT-Uplink-Scheduling} to guarantee transmission reliability and higher link rates. An adaptive modulation and coding scheme (MCS) \cite{WF-5G-2021} has been used with a set of modulations denoted by $\mathscr{M}$ corresponding to a set of discrete rates $\{\boldsymbol{\beta_m}\}_{m \in \mathscr{M}}$ and SINR thresholds $\{\boldsymbol{\gamma_m}\}_{m \in \mathscr{M}}$. The corresponding discrete rate function is denoted by $f(.)$ and plotted in Fig.~\ref{fig_g} with respect to the shannon's rate function.  

\subsection{The Benchmark (Upper-bound) Scheduler} 
\label{sec:Benchmark}
A {\bf \emph{Benchmark}} upper-bound uplink scheduler has been designed in this section for jointly scheduling IoT devices or 5G-NR users over each RB. Specifically, a joint upper-bound problem for the SC and MCS allocation on each RB has been formulated, while maintaining fairness among the user/devices across all cells. Note that a similar scheduler can be designed for the downlink. 

\vspace{0.2cm}
The variables for the proposed joint problem are defined as follows:
\begin{itemize}
	\item $\boldsymbol{P_{(i,b)}^{s}}$: transmission power used by device/user $(i,b)$ on SC $s$ with minimum and maximum transmitted power defined as $\boldsymbol{\epsilon_{P}}$ and $\boldsymbol{P_{max}}$, respectively.
		\item $\boldsymbol{\gamma_{(i,b)}}$: SINR seen by device/user $(i,b)$ on SC $s$ with minimum and maximum SINR defined as $\boldsymbol{\epsilon_{\gamma}}$ and $\boldsymbol{\gamma_{max}}$, respectively.
	\item $\boldsymbol{\lambda_{(i,b)}}$: the upper-bound link rate seen by device $(i,b)$ from gNodeB $b$ using $g(.)$ to envelop the underlying MCS (plotted in Fig.~\ref{fig_g}). Since, $g(.)$ is always greater than $f(.)$, $\boldsymbol{\lambda_{(i,b)}}$ will always be greater than or equal to the link rate obtained via $f(.)$.
	
\end{itemize}
Here, $\boldsymbol{\epsilon_P}$ and $\boldsymbol{\epsilon_{\gamma}}$ are very small positive values that depend on the numerical values of $\boldsymbol{P_{max}}$, and $\boldsymbol{\gamma_{max}}$, respectively. 

\vspace{0.2cm}
The optimization problem representing the benchmark upper-bound scheduling problem for time-slot $t$ of realization $\omega$ is defined as follows:
\begin{figure}
	\centering
	\includegraphics[width=9cm,height=5cm]{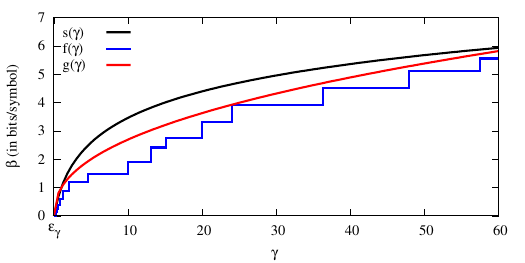}
	\vspace{-0.7cm}
	\caption{Shannon's rate function \big($s(\gamma)=\log_2(1+\gamma)$\big) vs. continuous rate function \big($g(\gamma)=\gamma^{\log_{10}(e)}$\big) vs. piece-wise discrete rate function \big($f(\gamma)$\big)}
	\label{fig_g}
	\vspace{-0.5cm}
\end{figure} 

\begin{equation*}
\boldsymbol{P_{Joint}^{UB}(\omega,t)}: 
\end{equation*}
\begin{equation*}
\underset{\{P_{(i,b)}^{s}\}, \{\gamma_{(i,b)}^{s}\}, \{\lambda_{(i,b)}\} }     {\text{maximize}}  \sum\limits_{ (i,b) \in (I_b(\omega),\mathscr{B})} \log{ ( \lambda_{(i,b)})}
\end{equation*}
\vspace{-0.5cm}
\begin{subequations}
	\begin{alignat}{3}
	&\text{subject to:}& \nonumber \\
	& \boldsymbol{\epsilon_P} \leq  P_{(i,b)}^{s} \leq   \boldsymbol{P_{max}},  & \forall (i,b), \forall s\\
	&\sum\limits_{s \in \mathscr{S}}  P_{i,b}^{s} \leq \boldsymbol{P_{max}}, \quad \quad  \quad  & \forall (i,b)& \\
	& \prod\limits_{s \in \mathscr{S}} P_{(i,b)}^{s} \leq {\boldsymbol{\epsilon_P}}^{(|\mathscr{S}|-1)}, & \forall (i,b) & \\
	& \prod\limits_{(i,b) \in (I_b(\omega),\mathscr{B})} P_{(i,b)}^{s} \leq {\boldsymbol{\epsilon_P}}^{(|I_b(\omega)|-1)}, & \forall s& \\
		& \boldsymbol{\epsilon_\gamma} \leq \gamma_{(i,b)}^{s} \leq   \boldsymbol{\gamma_{max}},  \qquad  \qquad  \qquad \quad \: & \forall (i,b),\forall s& 
	\end{alignat}
\end{subequations}
\begin{subequations}
	\begin{alignat}{3}
	& P_{(i,b)}^{s} G_{(i,b),b}(\omega,t) \geq  &\nonumber \\
	& \gamma_{(i,b)}^{s}  \big(  \boldsymbol{N_0} + \sum\limits_{(i',b') \in (I_b(\omega),\mathscr{B})}^{(i',b') \neq (i,b)} P_{(i',b')}^{s}  G_{(i',b'),b}(\omega,t) \big),  &\nonumber \\
	& \qquad \qquad \qquad \qquad \qquad \qquad \quad \quad \quad \quad \: \: \forall (i,b), \forall s& \tag{2f} \\
	&\lambda_{(i,b)}=\sum\limits_{s \in \mathscr{S}} ( \gamma_{(i,b)}^{s} )^{log_{10}(e)},   \qquad \qquad \quad \quad \: \: \forall (i,b) \tag{2g}& 
	\end{alignat}
\end{subequations}
Here, the constraints (2a) and (2b) are for assigning appropriate device power on each SC corresponding to the SINR variables in constraints (2e) and (2f). In the absence of binary variables, the multi-linear constraints (2c) and (2d) are required to ensure that a device/user does not transmit on more than one SCs and also to ensure that two or more devices/users do not transmit on the same SC. The continuous upper-bound function has been used in constraints (2g) to envelop the underlying MCS. 

The benchmark upper-bound problem $\boldsymbol{P_{Joint}^{UB}(\omega,t)}$ can be transformed into an equivalent convex problem $\boldsymbol{P_{Joint}^{UB'}(\omega,t)}$ by using the geometric programming (GP) transformations suggested by \cite{TCOM2018, TVT2019}. A logarithmic change of the variables and a logarithmic transformation of the objective function as well as the constraints is required for GP  transformations. For consistency, similar notations have been used to represent the transformed variables but with a prime symbol, i.e., $\boldsymbol{P_{(i,b)}^{'s}=\log(P_{(i,b)}^{s})}$ and $\boldsymbol{\gamma_{(i,b)}^{'s}=\log(\gamma_{(i,b)}^{s})}$. Further, $\boldsymbol{LSE(\{x\})}$ notation has been used to denote the log sum of exponentials over the set $\{x\}$. The objective function of the benchmark upper bound problem can be rewritten as a sum of $\boldsymbol{LSE}$ functions as follows:
\begin{align*}
		&\sum\limits_{ (i,b) \in (I_b(\omega),\mathscr{B})} \log{ (\lambda_{(i,b)} )} \nonumber\\
		&=\sum\limits_{ (i,b) \in (I_b(\omega),\mathscr{B})} \log{  (\sum\limits_{s \in \mathscr{S}}   {\gamma_{(i,b)}^{s}}^{log_{10}(e)})} \\
		&=\sum\limits_{ (i,b) \in (I_b(\omega),\mathscr{B})} \log{ ( \sum\limits_{s \in \mathscr{S}}   e^{\log({\gamma_{(i,b)}^{s}}^{log_{10}(e)})})}\\
		&=\sum\limits_{ (i,b) \in (I_b(\omega),\mathscr{B})} \log{ ( \sum\limits_{s \in \mathscr{S}}   e^{log_{10}(e)*\gamma_{(i,b)}^{s'}})}\\
		&=\sum\limits_{ (i,b) \in (I_b(\omega),\mathscr{B})} \boldsymbol{LSE}\big({\{ log_{10}(e)*\gamma_{(i,b)}^{s'}\}_{s \in \mathscr{S}}}\big)
\end{align*}

Note that the multi-linear constraints from (2f) can be converted into \emph{posynomial} functions that are non-convex in nature; however, a logarithmic transformation, as described in \cite{boyd2004convex}, can convert them into $\boldsymbol{LSE}$ functions in (3f), which are convex in nature. Further, the variables $\lambda_{(i,b)}$ and the corresponding constraints (2g) can been omitted from $\boldsymbol{P_{Joint}^{UB'}(\omega,t)}$, because they have been incorporated in the objective function. 

After using aforementioned GP transformations, we re-formulate the benchmark upper-bound problem as follows:

\setcounter{equation}{2}
\begin{equation*}
	\boldsymbol{P_{Joint}^{UB'}(\omega,t)}: 
\end{equation*}
\begin{equation*}
	\underset{\{P_{(i,b)}^{'s}\}, \{\gamma_{(i,b)}^{'s}\}, }     {\text{maximize}}  \sum\limits_{ (i,b) \in (I_b(\omega),\mathscr{B})} \boldsymbol{LSE}\big({\{ log_{10}(e)*\gamma_{(i,b)}^{s'}\}_{s \in \mathscr{S}}}\big)
\end{equation*}
\begin{subequations}
\begin{alignat}{3}
			&\text{subject to:}& \nonumber \\
		& \log(\boldsymbol{\epsilon_P}) \leq P_{(i,b)}^{'s}  \leq \log( \boldsymbol{P_{max}}),    \quad \quad  \quad \: \forall (i,b),  \forall s &\\
			& LSE({\{P_{(i,b)}^{'s}\}}_{s \in \mathscr{S}})  \leq \log(\boldsymbol{P_{max}}),   \quad  \qquad \quad \forall (i,b) & \\
				& \sum\limits_{s \in \mathscr{S}} P_{(i,b)}^{'s} \leq  {(|\mathscr{S}-1)} \times \log({\boldsymbol{\epsilon_P}}) ,  \quad \quad  \quad \quad \forall (i,b) & \\
			& \sum\limits_{(i,b) \in (I_b(\omega),\mathscr{B})} P_{(i,b)}^{'s} \leq  {(|I_b(\omega)|-1)} \times \log({\boldsymbol{\epsilon_P}}) ,  \forall s & 		
	\end{alignat}
\end{subequations}
\begin{subequations}
	\begin{alignat}{3}
	& \log(\boldsymbol{\epsilon_\gamma}) \leq \gamma_{(i,b)}^{'s} \leq   \log(\boldsymbol{\gamma_{max}}) ,   \quad  \quad  \qquad \quad \: \forall (i,b),\forall s& \tag{3e}\\
		&\log\bigg(  \exp\big( \gamma_{(i,b)}^{'s}- P_{(i,b)}^{'s} +
				\log(N_0) -\log(G_{(i,b),b}(\omega))\big) \nonumber\\
				&+\sum\limits_{(i',b') \in (I_b(\omega),\mathscr{B})}^{(i',b') \neq (i,b)} \exp\big( \gamma_{(i,b)}^{'s} + P_{(i',b')}^{'s}-  P_{(i,b)}^{'s} \nonumber  \\
				&+\log(G_{(i',b'),b}(\omega))-\log(G_{(i,b),b}(\omega)) \big) \bigg) \leq  0 ,  \forall (i,b) , \forall s \tag{3f} 
	\end{alignat}
\end{subequations}
Note that an optimal solution for $\boldsymbol{P_{Joint}^{UB}(\omega,t)}$ obtained by solving $\boldsymbol{P_{Joint}^{UB'}(\omega,t)}$ will provide an upper-bound link rate, i.e., $\lambda_{(i,b)}$, which will be higher than the one obtained via the discrete rate function $f(.)$. Hence, these upper-bound link rates will be converted into the corresponding discrete rates in Section~\ref{sec:Benchmark-results} by using $\boldsymbol{\gamma_{(i,b)}^{'s}}$ and the set of coding efficiencies $\{\boldsymbol{\beta_m}\}_{m \in \mathscr{M}}$ corresponding to the set of SINR thresholds $\{\boldsymbol{\gamma_m}\}_{m \in \mathscr{M}}$.

\subsection{The Baseline Scheduler}
\label{sec:Baseline}
In this section, a baseline scheduler has been defined for joint SC and MCS allocation in timeslot $t$ of realization $\omega$. The  {\bf \emph{Baseline}} scheduler allocates one SC to each device/user in a round-robin fashion using maximum available transmit power (i.e., $P_{(i,b)}^{s}(\omega, t)=\boldsymbol{P_{max}}, \forall \omega, \forall t$, when SC $s$ is allocated to device $(i,b)$ and $0$ otherwise). The same adaptive MCS for computing {\bf \emph{Estimated}} link rates has been used, i.e., $\tilde{\lambda}_{(i,b)}$, where SINR for each device/user is estimated using fixed ICI compensation. 

The {\bf\emph{Effective}} link rate ($\lambda_{(i,b)}$) will be lower than the {\bf \emph{Estimated}} one, because ICI compensation can lead to decoding errors at the gNodeB, thus would require re-transmissions (i.e., in the next 5G-NR frame) with {\bf \emph{MCS Adjustment}}. The $\lambda_{(i,b)}=0$ when the MCS assigned to $(i,b)$ (i.e., using the estimated SINR) is higher than the MCS with the effective SINR. This scenario can be corrected with {\bf \emph{MCS Adjustment}} in the next re-transmission by assigning MCS using the effective SINR from the first transmission. The baseline scheduler with re-transmission is named as {\bf\emph{Baseline-Rx-Tx}}.

%% file: Deep-Learning.tex
\section{Proposed DRL Framework}
\label{sec:DRL}
The proposed multi-agent DRL framework, for obtaining feasible solutions for the benchmark scheduling problem proposed in Section~\ref{sec:Benchmark}, consists of one learner/actor per user/device type (i.e., NB-IoT, LTE-M, and 5G-NR) within each gNodeB. 

A set of different DRL algorithms have been designed in this section, where each one of them determines how an agent takes a sequence of actions within a cellular environment to maximize a given reward function. The SC allocations are computed locally for each gNodeB whenever there is a change in the network; for simplicity, a round robin scheduler has been used. These allocations along with the channel state information is used by the learner for learning a policy to determine interference allocations. The learned policy is used locally by the corresponding actors to determine their power allocations and the corresponding MCSs. 

For centralized DRL learning, the learners reside within a BBU pool of a cloud-RAN (C-RAN) architecture; where, the C-RAN acts as a centralized learner for learning a scheduling policy and forwarding scheduling decisions to users/devices via their respective actors. While one may be tempted to co-locate the learner and actors within a BBU pool, this would require storing a large amount of data directly in the C-RAN hardware and losing the advantages of the actor-learner split. Further, the learners have to operate synchronously in order to achieve rewards that can be measured via inter-agent coordination. Nonetheless, the amount of memory, training time, and the size of DNN parameters required for centralized training with inter-agent coordination can be significantly reduced via edge learning. Further, retraining, and redeployment for continued prediction accuracy is faster than the centralized approach.

\subsection{DRL Preliminaries}
 The environment for the DRL framework is formulated as a discrete-time model-based Markov decision process (MDP) \cite{busoniu2017reinforcement} with a finite set of actions and states denoted by $A$ and $S$, respectively. Each agent interacting with the environment takes action $\alpha \in A$ by using policy $\pi$ for the observed state $\varsigma \in S$, and then receives a feedback reward $r$, and transitions to a new state $\varsigma' \in S$ within the environment. 
 
 The goal is to compute an optimal policy $\pi$ to maximize a reward function via continuous interactions with the environment without knowledge of the reward function and the state transitions. The general reward function for multi-agent systems at time $t$ (also known as the future cumulative discounted reward) is defined as follows: 
\setcounter{equation}{3}
\begin{equation}
R(t) = \sum_{\tau=0}^{\infty} d^{\tau}r(t+\tau+1)
\end{equation}
Here, $r(t+\tau+1)$ denotes the instantaneous reward at time $t+\tau+1$ that depends on current state and action (i.e., state-action pair). $d \in (0, 1]$ is the discount-rate parameter typically used to adjust the preference for incorporating future rewards either over a shorter ($d \rightarrow 0$) or a longer duration ($d \rightarrow 1$). 

In an urban cellular network, a network realization $\omega$ can change frequently leading to high variance and slow convergence of the DRL algorithms. Therefore, the hyper-parameters $d$ and $\tau$ are set to zero in (4), thereby the Q-function will depend on the expected value of the instantaneous reward only. 

\newpage
\subsection{DRL Assumptions and Design}
The action space ($A$) for all agents within the framework depends on the ICI for each user/device $(i,b)$. For a given realization $\omega$ and timeslot $t$, the actions are defined as follows:
\begin{equation}
\alpha_{(i,b)}(\omega, t):=\phi_{(i,b)}^{s_{(i,b)}} (\omega,t)
\end{equation}
Here, the interference level allocated to user/device $(i,b)$ on $s_{(i,b)}$ (i.e., the sub-carrier allocated to user/device $(i,b)$) is denoted by $\phi_{(i,b)}^{s_{(i,b)}} (\omega,t)$. Since, the action space ($A$) for DQN and PGN is finite, a set of discrete interference levels has been used that are uniformly spaced between $\boldsymbol{\Phi_{min}}$ and $\boldsymbol{\Phi_{max}}$. Note that interference levels can be discretized effectively due to their smaller range as compared to the power levels; further, the discretization of power does not work well with discrete rates (for a given MCS), since, SINR is more susceptible to quantization error in power as compared to interference. 

The power assigned to each user/device $(i,b)$ on $s_{(i,b)}$, which is denoted by $P_{(i,b)}^{s_{(i,b)}}(\omega,t)$, can be computed using $\phi_{(i,b)}^{s_{(i,b)}} (\omega,t)$ as follows: 
\begin{equation}
P_{(i,b)}^{s_{(i,b)}}(\omega,t)=     min \bigg(\boldsymbol{P_{max}},   \boldsymbol{\gamma_{max}}  \frac{   \boldsymbol{N_0} + \phi_{(i,b)}^{s_{(i,b)}} (\omega,t)} {{G}_{(i,b),b}(\omega,t) }\bigg)
\end{equation}
The instantaneous link rate during realization $\omega$ and timeslot $t$ on $s_{(i,b)}$ for each user/device $(i,b)$ using (6) is defined as follows:
\begin{equation}
\begin{split}
	\lambda_{(i,b)}^{s_{(i,b)}} (\omega, t) &:=  f ( \gamma_{(i,b)}^{s_{(i,b)}} (\omega, t) ) \\
	\gamma_{(i,b)}^{s_{(i,b)}} (\omega, t) ) &:= \frac{ {P}_{(i,b)}^{s_{(i,b)}} \times {G}_{(i,b),b}(\omega,t) }{N_{0} + \sum\limits_{(i',b') \in (I_b(\omega),\mathscr{B})}^{(i',b') \neq (i,b)} P_{(i',b')}^{s_{(i,b)}}  G_{(i',b'),b}(\omega,t))} 
\end{split}
\end{equation}
Using (6) and (7), the state of an agent $b$ ($\varsigma_b \in S$) when serving user/device $(i,b)$ is defined as follows:
\begin{equation}
\begin{split}
\varsigma_{(i,b)}(\omega, t) :=& \bigg\{ \Big\{ \frac { G_{(i',b'),b} (\omega,t) } { G_{(i,b),b} (\omega,t)} \Big\}_{ (i',b') \in I^{s_{(i,b)}}(\omega)},\\
&P_{(i,b)}^{s_{(i,b)}} (\omega,t-1), \lambda_{(i,b)}^{s_{(i,b)}} (\omega,t-1)\bigg\}
\end{split}
\end{equation}
Here, $I^{s_{(i,b)}}(\omega)$ denotes the set of users/devices from all cells that are transmitting on sub-carrier $s_{(i,b)}$ during realization $\omega$. Note that the exact SINR and the corresponding link rates and MCS for timeslot $t$ cannot be measured exactly beforehand without a C-RAN. 

The state of each agent depends on user/device $(i,b)$'s power and link rate in the previous state as well as the current channel-gains for all users/devices in $I^{s_{(i,b)}}(\omega)$. In contrast, the state of each agent in the centralized DRL setting depends on the current channel-gains as well as the power allocations and link rates in the previous state for all users/devices in $I^{s_{(i,b)}}(\omega)$, which requires inter-agent coordination after each timeslot or centralized training; assuming that BSs can measure interfering channel gains on their uplink sub-carriers and vice-versa on the downlink. 

For edge learning, the rewards for each agent $b$ in the proposed DRL framework are computed as follows:
\begin{equation}
	\begin{split}	  
		R_b(\omega,t+1)=
		\big\{\lambda_{(i,b)}^{s_{(i,b)}}(\omega, t)\big\}_{i \in I_b(\omega)}
	\end{split}
\end{equation}
For centralized training, the rewards are computed collectively using inter-agent coordination as follows:
\begin{equation}
	\begin{split}	  
		R(\omega,t+1)=
		\bigg\{\sum\limits_{(i',b')  \in I^{s_{(i,b)}}(\omega)} \lambda_{(i',b')}^{s_{(i,b)}}(\omega, t)\bigg\}_{(i,b) \in (I_b(\omega),\mathscr{B})}
	\end{split}
\end{equation}

%% file: Deep-Learning-Algorithms.tex
\subsection{Deep Q-Learning Network Design}
\label{sec:DRL-DQN}
A DQN algorithm typically learns an action-value function, namely a Q-function, to estimate the long term reward of a state-action pair. Thus, for a given policy $\pi$, the Q-function seen by an agent in state $\varsigma$ at time $t$, taking action $\alpha$, and thereafter following policy $\pi$, is defined as the expected sum of the future cumulative discounted rewards as follows: 
\begin{equation}
	Q_\pi(\varsigma,\alpha) : = E_\pi \big[   R(t) \big|  \varsigma(t)= \varsigma, \alpha(t)=\alpha\big]
\end{equation}
Here, $E_\pi[.]$, the expectation operator for maximizing the long-term reward, is equivalent to searching a policy that maximizes the Q-function. When a DQN is modeled as an MDP, the optimal Q-function satisfies the Bellman optimality equation, thus, it can sample and learn the Q-function by using the Bellman equation as an iterative update \cite{sutton2018reinforcement}. 

The simplest approach to Q-learning is to store and update Q-values for individual state-action pairs within a table. While, tabular-based Q-learning has also been suggested for wireless applications, it is unsuitable for RRM mainly because of the large dimensionality of the underlying state-action space. To mitigate this issue, it has been suggested in \cite{calabrese2018learning} to apply Q-learning via functional approximation of the Q-function. Q-learning when combined with functional approximation can speed up learning in finite state problems. Since, Q-learning based algorithms can generalize earlier experiences into previously unseen states, this combination makes it possible to apply DQN algorithms to larger RRM problems, even when the state space is continuous.

Although different functional approximators could be used within the learning framework, an adapted artificial neural network (ANN) has been selected due to its superb generalization capabilities and the existence of computationally efficient training algorithms \cite{goodfellow2016deep}. To train the ANN given a set of samples of experience $\{(\varsigma(t), \alpha(t), r(t+1), \varsigma(t+1))\}$, an iterative approach has been used to reconstruct a training set consisting of state-action pairs $(\varsigma(t) ,\alpha(t))$ as input.

The DQN algorithm (outlined in Algorithm~\ref{DQN-Algo}) describes the design of the DQN, i.e., denoted by $Q(\varsigma,\alpha;\boldsymbol{w^q_{b}})$, where $\boldsymbol{w^q_{b}}$ denotes the set of optimization parameters to minimize the mismatch between current Q-value and the target Q-value. An optimal action $\alpha^{*} \in A$ within a DQN tends to maximize the $Q$ value (i.e.,  $\underset{\alpha}{\text{ argmax }} Q(\varsigma,\alpha; \boldsymbol{w^q_{b}})$). The selection of a good action (i.e., $\alpha$) relies on accurate estimation of $\boldsymbol{w^q_{b}}$, therefore, an underlying DNN ($l_q=$ learning rate) is required to search for the optimal parameters for $\omega$ as follows: 
\begin{equation}
	\begin{split}
		\boldsymbol{w_{b}^{q*}} = &\underset{\boldsymbol{w^q_{b}}}{\text{ argmin }}  \frac{1}{2} {\big( Q(\varsigma,\alpha;\boldsymbol{w^q_{b}}) - R_{b}(\omega,t+1) \big)}^2 \\
		\nabla \boldsymbol{w^q_{b}} = &\big(Q(\varsigma, \alpha; \boldsymbol{w^q_{b}})- R_{b}(\omega,t+1) \big) \nabla_{\boldsymbol{w^q_{b}}} Q(\varsigma, \alpha;\boldsymbol{w^q_{b}})
	\end{split}
\end{equation}
Here, $\boldsymbol{w_{b}^{q*}}$ denotes the set of optimal parameters for the DQN and $\nabla \boldsymbol{w^q_{b}}$ is the gradient vector for minimizing the mean squared loss function.

The DQN algorithm adopts an $\epsilon$-greedy policy to control the exploration probability, i.e., $\epsilon$, where each agent uses both exploration of new possibilities and exploitation of prior knowledge. When $\epsilon$ is set to 0, the algorithm never explores but always exploit the knowledge it already has. On the contrary, having the epsilon set to 1 forces the algorithm to always take random actions and never use past knowledge. 

DNN algorithms are prone to over-fitting and it is hard to produce meaningful sample of experiences, therefore, \emph{Experience Replay} has been exploited in Algorithm~\ref{DQN-Algo} to store experiences including state transitions, rewards and actions. Further, a  set of mini-batches have been used to update the underlying DNN, which has the following merits: it reduces correlation between experiences when updating the DNN, it increases learning speed with mini-batches, and this technique reuses past transitions to avoid forgetting past learning experiences.

\begin{algorithm}[t]
	\caption{DQN: Edge Learning Algorithm}\label{DQN-Algo}
	\begin{algorithmic}
			\For{each agent $b \in \mathscr{B}$}{}
		
		\State Initialize $Q(\varsigma,\alpha; \boldsymbol{w^q_b})$ with
		random parameters $\boldsymbol{w^q_b}$.
		\State Initialize Replay Memory $D_{b}$ with size $B_D$.
		\State Initialize Sum of Rewards $R_{b},R^*_b=0$.

		\EndFor
		\For{each $\omega \in \Omega$}{}
		\For{each agent $b \in \mathscr{B}$}{}
		
		\State Determine $\{s_{(i,b)}\}_{i \in I_b(\omega)}$.
		\State Set  $\{\phi_{(i,b)}^{s_{(i,b)}}(\omega,0)\}_{i \in I_b(\omega)}=\{0\}$. 
		\State Set  $\{P_{(i,b)}^{s_{(i,b)}}(\omega,0)\}_{i \in I_b(\omega)}=\{0\}$.
		\State Set initial state $\{\varsigma_{(i,b)}^a(\omega,1)\}_{i \in I_b(\omega)}$.
		\State Set initial state $\{\varsigma_{(i,b)}^c(\omega,1)\}_{i \in I_b(\omega)}$.

		\For{each $t \in \mathscr{T}$}{}
		\State Select $\epsilon \in [0,1]$ randomly.
        
	\For{each agent $i \in I_b(\omega)$}{}
		\If{$\epsilon < \epsilon_q$} {}
		
		\State Choose $\alpha_{(i,b)}(\omega, t)$ randomly.
		
		\Else
		\State Set $\alpha_{(i,b)}(\omega,t)=$
		\State $\underset{\alpha}{\text{ argmax }} Q(\varsigma_{(i,b)}, \alpha; \boldsymbol{w^q_{b}})$.

		\EndIf
		
		\State Compute  
		$\phi_{(i,b)}^{s_{(i,b)}}(\omega,t)$ using (5).
		\State Compute $P_{(i,b)}^{s_{(i,b)}}(\omega,t)$ using (6).
		\State Compute $\lambda_{(i,b)}^{s_{(i,b)}} (\omega, t)$ using (7).
		\State Set  $r_{(i,b)}(\omega, t+1)=\lambda_{(i,b)}^{s_{(i,b)}} (\omega, t)$
		
		\State Store $\{\varsigma_{(i,b)}(\omega,t), \alpha_{(i,b)}(\omega,t),$ 
		\State $r_{(i,b)}(\omega,t+1),\varsigma_{(i,b)}(\omega,t+1)\}$ in $D_{b}$.
		\State Update $\varsigma_{(i,b)}(\omega, t) \leftarrow \varsigma_{(i,b)}(\omega,t+1)$
		\EndFor
		
		\EndFor

		\If{ atleast $N_D$ experiences in $D_{b}$} {}
		\State Sample $N_D$ experiences from $D_{b}$. 
		\State Compute $\boldsymbol{w^q_{b}}$ that minimizes loss using (12).
		\State Compute $R_{b}=$ sum of all $N_D$ rewards 
		\If{ $R_{b} > R^*_b $ }
		\State Update $\boldsymbol{w^{q*}_b} \leftarrow \boldsymbol{w^q_{b}}$
		\State Update $R^*_b \leftarrow R_{b} $
		
		\EndIf 
		\EndIf    
		\EndFor
		
		\EndFor
		\State Output: Trained DQN  $Q(\varsigma,\alpha;\boldsymbol{w^{q*}_b})$.
	\end{algorithmic}

\end{algorithm}
\begin{algorithm}[t]
	\caption{PGN: Edge Learning Algorithm}\label{PGN-Algo}
	\begin{algorithmic}
			\For{each agent $b \in \mathscr{B}$}{}
			
		\State Initialize PGN $\pi(\alpha|\varsigma; \boldsymbol{w^\pi_{b}})$ with random $\boldsymbol{w^\pi_{b}}$.
			\State Initialize Reward Memory $D_{b}$ with size $N_D$.
		\State Initialize Sum of Rewards $R_{b},R^*_b=0$.

		\EndFor
		\For{each $\omega \in \Omega$}{}
		\For{each agent $b \in \mathscr{B}$}{}
		\State Determine $\{s_{(i,b)}\}_{i \in I_b(\omega)}$ .
	\State Set  $\{\phi_{(i,b)}^{s_{(i,b)}}(\omega,0)\}_{i \in I_b(\omega)}=\{0\}$. 
	\State Set  $\{P_{(i,b)}^{s_{(i,b)}}(\omega,0)\}_{i \in I_b(\omega)}=\{0\}$.
	\State Set initial state $\{\varsigma_{(i,b)}(\omega,1)\}_{i \in I_b(\omega)}$.

		\For{each $t \in \mathscr{T}$}{}

			  \State Set $\{\alpha_{(i,b)}(\omega, t)\}_{i \in I_b(\omega)}$ following
			 \State  $\pi(\alpha|\varsigma_{(i,b)}; \boldsymbol{w^\pi_{b}})$ to find 
			 $\{\phi_{(i,b)}^{s_{(i,b)}}(\omega,t)\}_{i \in I_b(\omega)}$.
			 \State Compute $\{P_{(i,b)}^{s_{(i,b)}}(\omega,t)\}_{i \in I_b(\omega)}$ using (6).
			 \State Update $\varsigma_{(i,b)}^a(\omega,t) \leftarrow \varsigma_{(i,b)}^a(\omega,t+1)$
			 \State Update $\varsigma_{(i,b)}^c(\omega,t) \leftarrow \varsigma_{(i,b)}^c(\omega,t+1)$
			 \State Compute $\{\lambda_{(i,b)}^{s_{(i,b)}} (\omega, t)\}_{i \in I_b(\omega)}$ using (7).
			 \State Set  $\{r_{(i,b)}(\omega, t+1)\}_{i \in I_b(\omega)}$ 
			 
			 \State $=\{\lambda_{(i,b)}^{s_{(i,b)}} (\omega, t)\}_{i \in I_b(\omega)}$ and  Store in $D_b$

			 \State Set $R_{b}(\omega,t+1)=\{\lambda_{(i,b)}^{s_{(i,b)}}(\omega, t)\}_{i \in I_b(\omega)}$
			 
			 \State Find $\boldsymbol{w^{\pi}_b}$ that  maximizes $R_{b}(\omega,t+1)$ using (13).

	  	    \EndFor

		  \If{ $N_D$ rewards in $D_{b}$} {}
		  \State Compute $R_{b}=$ sum of all rewards in $D_{b}$
		  \If{ $R^*_{b} > R_{b} $ }
		  \State Update $\boldsymbol{w^{\pi *}_b} \leftarrow \boldsymbol{w^{\pi}_b}$
		  \State Update $R^*_b \leftarrow R_{b} $
		  \State Update $D_{b} \leftarrow \emptyset$
		  
		  \EndIf 
		  \EndIf      
		  
	  	  \EndFor
	  	   \EndFor
		\State Output: Learned PGN $\pi(\alpha|\varsigma; \boldsymbol{w^{\pi *}_b})$.
    \end{algorithmic}
 
\end{algorithm}
\subsection{Policy Gradient Network Design}
\label{sec:DRL-PGN}
The DQN described in Section~\ref{sec:DRL-DQN} learns the action-value function to select actions accordingly. In contrast, a PGN (i.e., policy gradient network) learns a stochastic policy $\pi$ by using a parameterized function, i.e., denoted by $\pi(\alpha|\varsigma; \boldsymbol{w^\pi_{b}})$. Generally, a PGN learning algorithm is derived using a Monte-Carlo policy gradient learning algorithm \cite{thomas2017policy}, which proceeds by sampling the stochastic policy $\pi$ and adjusting the set of policy parameters $\boldsymbol{w^\pi_{b}}$ in the direction of the greater reward. The PGN algorithm (outlined in Algorithm~\ref{PGN-Algo}) aims at maximizing the expected value of instantaneous reward as opposed to minimizing the mean squared loss for a DQN. 

The PGN is designed to find optimal policy parameters for agent $b$ in $\omega$ using the policy gradient theorem as follows:

 \begin{equation}
	\begin{split}
		\boldsymbol{w_{b}^{\pi*}} &= \underset{\boldsymbol{w^\pi_{b}}}{\text{ argmax }} E_\pi \bigg[   \sum\limits_{\alpha} \pi(\alpha|\varsigma; \boldsymbol{w^\pi_{b}}) R_{b}(\omega,t+1) \bigg] \\
		\nabla \boldsymbol{w^\pi_{b}} &=E_\pi \bigg[ R_{b}(\omega,t+1) \nabla_{\boldsymbol{w^{\pi}_{b}}} \ln \pi(\alpha|\varsigma; \boldsymbol{w_{b}^\pi})  \bigg]
	\end{split}
\end{equation}
Here, $\boldsymbol{w_{b}^{\pi*}}$ and $\nabla \boldsymbol{w^\pi_{b}}$  denote the set of optimal parameters for the PGN and the corresponding gradient vector for maximization, respectively.
\subsection{Deep Deterministic Policy Gradient Network Design} 
\label{sec:DRL-DDPG}
The design of an actor-critic algorithm (i.e., DDPGN) operating over a continuous action space has been discussed in this section, where an action-value function is learned by each agent in addition to the policy, while the critic updates the action-value function parameters $\boldsymbol{w^c_{(i,b)}}$ and the actor updates policy parameters $\boldsymbol{w^a_{(i,b)}}$ in the direction suggested by the critic. When the optimal policy in an actor-critic based network is deterministic, which is the usual case in most RRM problems \cite{sutton2018reinforcement}, it becomes equivalent to a DDPGN \cite{lillicrap2015continuous}. 

The DDPGN Training algorithm  described in Algorithm~\ref{DDPGN-Algo}, learns the deterministic policy in an off-policy manner, where both critic and actor work together to optimize their own parameters. The goal of the actor is to generate a deterministic action $\alpha$ using a deterministic policy network (i.e., denoted by $\pi_A(\varsigma^a; \boldsymbol{w^a_b})$ with $\varsigma^a$ being the actor's state from (8)). The critic can then predict the $Q$-value for the action-state pair through a critic network (i.e., denoted by $Q_C(\varsigma^c, \alpha; \boldsymbol{w^c_{b}})$) with $\varsigma^c$ being the critic's state, which is defined as follows:

\begin{equation}
	\begin{split}
			\varsigma^c_{(i,b)}(\omega, t) :=& \bigg\{ \{  G_{(i',b'),b} (\omega,t) \}_{ (i',b')  \in I^{s_{(i,b)}}(\omega)}\bigg\}
	\end{split}
\end{equation}
\newpage
The DDPGN algorithm aims for minimizing the critic's loss and maximizing actor's gain by applying gradient ascent (with actor's and critic's learning rates denoted by $l_a$ and $l_c$, respectively) for agent $b$ during $\omega$ as follows: 
\begin{equation}
\begin{split}
&\boldsymbol{w^{a*}_b} = \underset{\boldsymbol{w^a_b}}{\text{ argmax }} Q_C(\varsigma^c,\alpha=\pi_A(\varsigma^a_b; \boldsymbol{w^a_b});\boldsymbol{w^c_b})\\
&\boldsymbol{w^{c*}_b} = \underset{\boldsymbol{w^c_b}}{\text{ argmin }}  \frac{1}{2} \big( Q_C (\varsigma^c, \alpha=\pi_A(\varsigma^a_b; \boldsymbol{w^a_b});\boldsymbol{w^c_b}) \\		
	& \qquad \qquad - R_b(\omega,t+1) \big)^2  \\	
 &\nabla \boldsymbol{w^c_b} =\big(Q_C(\varsigma^c,\alpha=\pi_A(\varsigma^a_b; \boldsymbol{w^a_b}); \boldsymbol{w^c_b})- R_b(\omega,t+1) \big) \\
& \qquad \qquad \nabla_{\boldsymbol{w^c_b}} Q_C(\varsigma^c, \alpha=\pi_A(\varsigma^a_b; \boldsymbol{w^a_b});\boldsymbol{w^c_b})\\ 
&\nabla \boldsymbol{w^a_b}=\nabla_{\boldsymbol{\alpha}}  Q_C(\varsigma^c, \alpha=\pi_A(\varsigma^a_b; \boldsymbol{w^a_b}); \boldsymbol{w^c_b})\\ 
&\qquad \qquad \nabla_{\boldsymbol{w^a_b}}  \pi_A(\varsigma^a_b; \boldsymbol{w^a_b})
\end{split}
\end{equation}
Here, $\boldsymbol{w^{a*}_b}$ and $\boldsymbol{w^{c*}_b}$ denote the set of optimal parameters for the actor and critic, respectively. $\nabla \boldsymbol{w^a_b}$ and $\nabla \boldsymbol{w^c_b}$ are the corresponding gradient vectors. 

Since, the action space in DDPGN is continuous, it can be assumed that the $Q_C$ function is differentiable with respect to the set of actions. Thus, a gradient ascent can be executed (with respect to the policy parameters $\boldsymbol{w^a_b}$ only) to optimize actor's parameters by considering the critic's parameters ($\boldsymbol{w^c_b}$) as constants.
\begin{algorithm}[t!]
	\caption{DDPGN: Edge Learning Algorithm}\label{DDPGN-Algo}
	\begin{algorithmic}
		
		\For{each agent $b \in \mathscr{B}$}{}
		
		\State Initialize $\pi_A(\varsigma^a; \boldsymbol{w^a_b})$ and $Q_C(\varsigma^c,\alpha; \boldsymbol{w^c_b})$ 
		\State with
		random parameters $\boldsymbol{w^a_b}$ and $\boldsymbol{w^c_b}$.
		\State Initialize Reward Memory $D_{b}$ with size $N_D$.
		\State Initialize Sum of Rewards $R_{b},R^*_b=0$.
		
		\EndFor

		\For{each $\omega \in \Omega$}{}
		\For{each agent $b \in \mathscr{B}$}{}
		
		\State Determine $\{s_{(i,b)}\}_{i \in I_b(\omega)}$ .
		\State Set  $\{\phi_{(i,b)}^{s_{(i,b)}}(\omega,0)\}_{i \in I_b(\omega)}=\{0\}$. 
		\State Set  $\{P_{(i,b)}^{s_{(i,b)}}(\omega,0)\}_{i \in I_b(\omega)}=\{0\}$.
		\State Set initial state $\{\varsigma_{(i,b)}^a(\omega,1)\}_{i \in I_b(\omega)}$.
		\State Set initial state $\{\varsigma_{(i,b)}^c(\omega,1)\}_{i \in I_b(\omega)}$.

		\For{each $t \in \mathscr{T}$}{}
		
		\State Set $\{\alpha_{(i,b)}(\omega, t)\}_{i \in I_b(\omega)}$ following
		\State  $\pi_A(\varsigma^a_{(i,b)}; \boldsymbol{w^a_b})$ to find 
		$\{\phi_{(i,b)}^{s_{(i,b)}}(\omega,t)\}_{i \in I_b(\omega)}$.
		\State Compute $\{P_{(i,b)}^{s_{(i,b)}}(\omega,t)\}_{i \in I_b(\omega)}$ using (6).
		\State Update $\varsigma_{(i,b)}^a(\omega,t) \leftarrow \varsigma_{(i,b)}^a(\omega,t+1)$
		\State Update $\varsigma_{(i,b)}^c(\omega,t) \leftarrow \varsigma_{(i,b)}^c(\omega,t+1)$
		\State Compute $\{\lambda_{(i,b)}^{s_{(i,b)}} (\omega, t)\}_{i \in I_b(\omega)}$ using (7).
		\State Set  $\{r_{(i,b)}(\omega, t+1)\}_{i \in I_b(\omega)}$ 
		
		\State $=\{\lambda_{(i,b)}^{s_{(i,b)}} (\omega, t)\}_{i \in I_b(\omega)}$ and  Store in $D_b$

		\State Set $R_{b}(\omega,t+1)=\{\lambda_{(i,b)}^{s_{(i,b)}}(\omega, t)\}_{i \in I_b(\omega)}$
		
		\State Compute $\boldsymbol{w^{c}_b}$ and $\nabla \boldsymbol{w^a_b}$ using (15).
		\State Update $ \boldsymbol{w^a_b} \leftarrow  \boldsymbol{w^a_b} + l_a \nabla \boldsymbol{w^a_b}$

		\EndFor		
		\If{$N_D$ rewards in $D_{b}$} {}
		\State Compute $R_{b}=$ sum of all rewards in $D_{b}$
		\If{$R^*_{b} > R_{b} $ }
		\State Update $\boldsymbol{w^{a*}_b} \leftarrow \boldsymbol{w^a_{b}}$
		\State Update $\boldsymbol{w^{c*}_b} \leftarrow \boldsymbol{w^c_{b}}$
		\State Update $R^*_b \leftarrow R_{b} $
		\State Update $D_{b} \leftarrow \emptyset$
		
		\EndIf 
		\EndIf      
		
		\EndFor

		\EndFor
		\State Output: Learned $\pi_A(\varsigma^\alpha; \boldsymbol{w^{a*}_b})$ and $Q_C(\varsigma^c,\alpha; \boldsymbol{w^{c*}_b})$.
	\end{algorithmic}
\end{algorithm}

%% file: Results.tex
\section{Numerical Simulation Setup}
\label{sec:setup}
The numerical simulations are performed on a cellular network with $7$ adjacent cells (i.e., 3 sectors per cell site with reuse 3 and Inter-site Distance (ISD) of 500m). For simplicity, a wrapped around hexagonal multi-cell environment\footnote{The proposed framework can be used for different multi-cell environments.} has been used, where a gNodeB is placed at the center of each cell with a set of uniformly distributed NB-IoT devices, LTE-M devices, and 5G-NR UEs. The physical layer parameters for obtaining numerical results are summarized in Table \ref{tab:phy}. The transmission bandwidth for both NB-IoT and LTE-M RBs is set to 200kHz with 12 15kHz SCs in each RB. Although IoT devices can support multiple carrier operation, the numerical analysis is based on a single RB. Since, a single RB can support more than 52,500 IoT devices per cell \cite{3GPP-TR-45.820}, as not all of these devices will be active at the same time. Also, allocating more SCs (or higher bandwidth) to a power limited IoT device could lead to spectral inefficiency. 

For performance evaluation, a set of realizations ($\Omega$) has been used, where the timeslots per realization are set to $T = 20$; a large value of $|\Omega|$ and a small $T$ has been chosen to reduce the data correlation and the convergence time for the DRL algorithms. For each realization $\omega$, there are $N=84$ 5G-NR users or IoT devices located uniformly across all cells. Their SC channel gains, i.e., introduced in Section \ref{sec:rate-model} depend on distance, path loss model, penetration loss, large fading gain, small scale fading gain, and antenna gains. A Rayleigh distributed large-scale fading component has been assumed for numerical simulations that takes both geometric attenuation and shadow fading into account, which is assumed to be invariant over each realization. The large scale fading coefficients are modeled by a log-normal distribution with zero mean and $10$dB standard deviation. Using Jake's model, the small-scale flat fading is modeled as a first-order complex Gauss-Markov process with the correlation coefficient $\rho$ determined by $\rho = J_0(2\pi f_d T_f)$, where $J_0(.)$ is the first kind zero-order Bessel function, $f_d$ is the maximum Doppler frequency, and $T_f$ is the time interval between adjacent realizations. The Doppler frequency $f_d=10 Hz$ and time period $T_f=10ms$ (i.e., frame length in 5G-NR with $T=20$ timeslots of duration $T_S=0.5ms$ each with $N_{S}=14$ symbols per timeslot) are adopted to simulate the small scale fading effects.

\begin{table*}[t!]
	\caption {Physical Layer Parameters from \cite{3GPP-TR-45.820},  \cite{3GPP-TR-36.802}, and \cite{3GPP-TS-36.213}. }  \label{tab:phy}
	\vspace{-0.3cm}
	\begin{center}
		\begin{tabular}{|c|c||c|c||c|c|}
			\hline
			RB Bandwidth & 180kHz  & SC spacing & 15kHz & SCs per RB & 12 \\
			\hline
			Noise Power & -174dBm/Hz & 	gNodeB Noise Figure & 5dB &  Penetration Loss & 20dB \\
			\hline
			gNodeB Antenna Gain  & 15dBi  &  NB-IoT/LTE-M/5G-NR UE Antenna Gain & 0dBi & $\boldsymbol{P_{max}}$& $23dBm$  \\
			\hline
			gNodeB Antenna Type  & Tri-directional  &  NB-IoT/LTE-M/5G-NR UE Antenna Type & Omni &  $\boldsymbol{P_{min}}$ & $-40dBm$   \\
			\hline	
			Path Loss (dB) & \multicolumn{2}{c||}{$128.1 + 37.6$ log$_{10}(d/1000), d \ge 35m$} & 	gNodeB  Directivity Gain & \multicolumn{2}{c|}{{min$(12(\frac{\theta}{65^o})^2, 20)$dB}} \\
			\hline
			MCS for NB-IoT & \multicolumn{2}{c||}{ $|\mathscr{M}|=6$ and 	$\boldsymbol{\beta_{max}}=1.18$ bits/symbol} & \multicolumn{3}{c|}{NB-IoT: $\boldsymbol{\Phi_{max}}=-95dBm$ and $\boldsymbol{\Phi_{min}}-100dBm$} \\ 
			\hline
			MCS for LTE-M & \multicolumn{2}{c||}{$|\mathscr{M}|=9$ and $\boldsymbol{\beta_{max}}=2.41$ bits/symbol}  &  \multicolumn{3}{c|}{LTE-M: $\boldsymbol{\Phi_{max}}=-96dBm$ and $\boldsymbol{\Phi_{min}}=-101dBm$}    \\
			\hline
			MCS for 5G-NR & \multicolumn{2}{c||}{ $|\mathscr{M}|=15$ and $\boldsymbol{\beta_{max}}=5.55$ bits/symbol}  & \multicolumn{3}{c|}{5G-NR: $\boldsymbol{\Phi_{max}}=-97dBm$ and $\boldsymbol{\Phi_{min}}=-102dBm$}     \\
			\hline
		\end{tabular}
	
	\end{center}
\end{table*}
\begin{table*}
	\caption {DRL Parameters }  \label{tab:drl}
	\vspace{-0.3cm}
	\begin{center}
		\begin{tabular}{|c||c|c|c|c||c|c|}
			\hline
			DQN parameters & $\epsilon_q=0.2$ &  $|A|=10$ & $l_q=1e-4$ & 	Experience Replay Memory & $N_D=500$ & $B_D=500K$ \\
			\hline
			PGN parameters & $N_D=500$  &  $|A|=10$ & $l_p=1e-4$ & DDPGN parameters & $N_D=500$  & $l_a=l_c=1e-4$ \\
			\hline
		\end{tabular}
		
	\end{center}
\end{table*}
\subsection{DRL Setting}
The results of the proposed DRL algorithms are obtained on a computing platform consisting of an Intel i7-7700HQ CPU and a Nvidia GTX-1050Ti GPU. The list of DRL parameters used for generating numerical results are summarized in Table \ref{tab:drl}, where a small value for the learning rate has been chosen deliberately, which is a hyper-parameter that controls change in the model each time the optimization parameters are updated. The architecture of the underlying DNN is designed to provide a regressive model with the least number of layers and neurons thereby reducing the computational complexity.
For all DRL algorithms (both centralized and edge learning), only two hidden layers (with 64 and 128 neurons) have been used that are coated with a rectified linear unit (ReLU) activation function, which was chosen for its ability to become monotonic with its derivative and also to deal with non-linearity within the layers. The Adam algorithm has been adopted as the optimizer for all DRL algorithms with $l2$ loss function. The Input and output layers are defined as a linear activation function with dimensions corresponding to the dimensions of the states and actions, respectively.

\subsection{Performance Metrics}
For proposed interference allocation based DRL algorithms from Section~\ref{sec:DRL}, the action space ($A$) for both DQN and PGN is based on a set of discrete interference levels that are uniformly spaced between $\boldsymbol{\Phi_{min}}$ and $\boldsymbol{\Phi_{max}}$. In contrast, for comparison with power allocation based DRL algorithms, the action space ($A$) for both DQN and PGN has been set to discrete power levels that are uniformly spaced between $\boldsymbol{P_{min}}$ and $\boldsymbol{P_{max}}$. 
For analyzing the performance of NB-IoT, LTE-M, and 5G-NR users/devices, the arithmetic mean ($\mathcal{AM}$), geometric mean ($\mathcal{GM}$), and harmonic mean ($\mathcal{HM}$) throughput for each $\omega \in \Omega$ is computed as follows:
\begin{equation}
\begin{split}
\mathcal{AM}(\omega)	& :=     \frac{1}{ T . N}  \sum\limits_{\forall t  }    \sum\limits_{  \forall (i,b)} \bigg( \frac{ \lambda_{(i,b)} (\omega,t) . N_S}{T_S  }\bigg) \\
\mathcal{GM}(\omega) 	& := \frac{1}{ T } \sum\limits_{\forall t  }  \sqrt[N]{  \prod\limits_{  \forall (i,b) }  \bigg(   \frac{ \lambda_{(i,b)}(\omega,t). N_S }{T_S} \bigg) } \\
\mathcal{HM}(\omega) 	& :=    \frac{1}{ T } \sum\limits_{\forall t  }  \frac{N}{\sum\limits_{  \forall (i,b)} \bigg( \frac{T_S}{\lambda_{(i,b)} (\omega,t) . N_S}\bigg)}
\end{split}
\end{equation}

Here, $\lambda_{(i,b)} (\omega,t)$ is the actual link rate seen by each user/device $(i,b)$ during realization $\omega$ and timeslot $t$, which can be computed using the discrete rate function $f(.)$ (from Fig.~\ref{fig_g}). Unlike $\mathcal{AM}$, $\mathcal{HM}$ gives less significance to high-value outliers, thus, providing a more accurate picture of the average link rates for all users/devices in the network. Further, maximizing $\mathcal{HM}$ is equivalent to minimizing average delay per timeslot, which is critical for devices that are sending time-sensitive information. In contrast, $\mathcal{GM}$ provides a measure of throughput fairness among the users/devices in the network.
\vspace{-0.1cm}
\subsection{Benchmark (Upper-bound)} 
\label{sec:Benchmark-results}
The upper-bound solutions from the benchmark scheduler in Section~\ref{sec:Benchmark} (i.e., denoted by {\bf \emph{Benchmark-$g(\gamma)$}}) are obtained by solving $\boldsymbol{P_{Joint}^{UB'}(\omega, t)}$ for all realizations (i.e., $\omega \in \Omega$) using a commercial solver (SNOPT \cite{snopt}), where we denote the corresponding solutions obtained via the underlying MCS (plotted in Fig.~\ref{fig_g}) as {\bf \emph{Benchmark-$f(\gamma)$}}. The corresponding average $\mathcal{AM}$, $\mathcal{GM}$, and $\mathcal{HM}$ throughputs have been plotted in Fig.~\ref{No-fading}. Note that without small scale fading, the feasible solutions are invariant over all timeslots of each realization $\omega$.   
\subsection{Baseline} 
\begin{figure*}[t!]
	\centering
	\begin{tabular}{cc}
		\includegraphics[width=9cm,height=4.8cm]{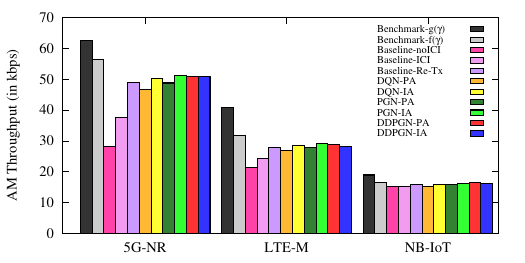}  & 
		\includegraphics[width=9cm,height=4.8cm]{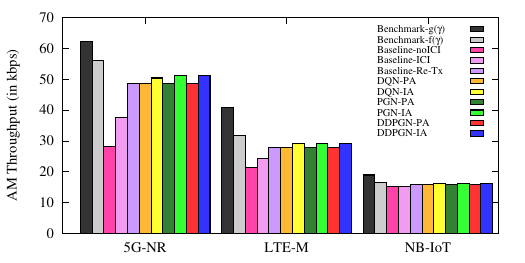} \\ 
		(a) $\Omega_{Train}$: $\mathcal{AM}$ \textit{Throughput}    &   (b) $\Omega_{Test}$: $\mathcal{AM}$ \textit{Throughput}  \\
		\includegraphics[width=9cm,height=4.8cm]{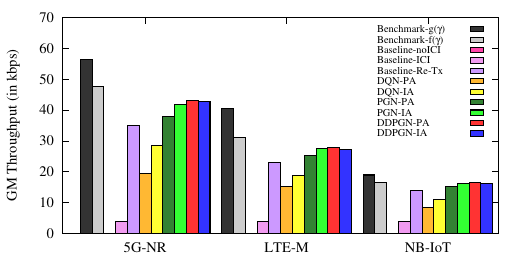}  & 
		\includegraphics[width=9cm,height=4.8cm]{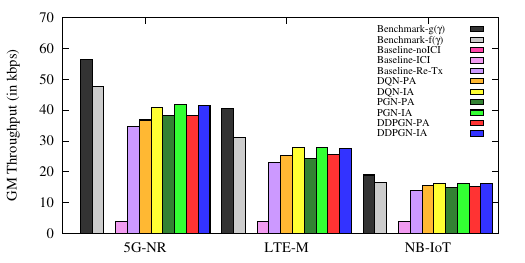} \\	
		(c) $\Omega_{Train}$: $\mathcal{GM}$ \textit{Throughput}  & (d) $\Omega_{Test}$: $\mathcal{GM}$ \textit{Throughput} \\
		\includegraphics[width=9cm,height=4.8cm]{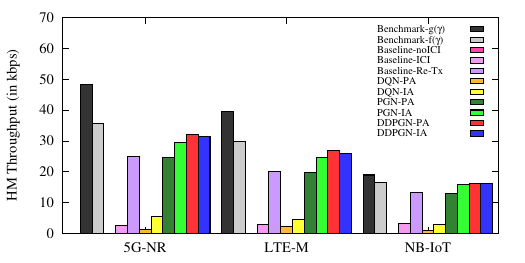}  & 
		\includegraphics[width=9cm,height=4.8cm]{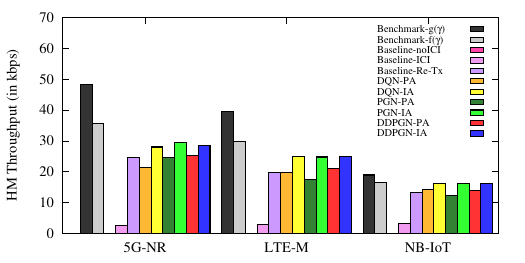} \\ 
		(e) $\Omega_{Train}$: $\mathcal{HM}$ \textit{Throughput}   &   (f) $\Omega_{Test}$: $\mathcal{HM}$ \textit{Throughput}  \\
		
	\end{tabular}
	\centering
	\caption{Centralized Training and Testing without small scale fading}
	\label{No-fading}
	
\end{figure*}

The baseline scheduling solutions from Section~\ref{sec:Baseline} for all realizations (i.e., $\omega \in \Omega$) are computed with and without taking ICI into consideration, i.e., denoted by {\bf \emph{Baseline-ICI}} and {\bf \emph{Baseline-noICI}}, respectively; where, different fixed ICI compensation factors have been used to individually maximize the average of $\mathcal{AM}$, $\mathcal{GM}$, and $\mathcal{HM}$ throughput. Note that with no ICI compensation some of the users/devices receive zero link rate leading to zero $\mathcal{GM}$ and $\mathcal{HM}$ (or infinite delay) in Fig.~\ref{No-fading} (c) to (f). The improvement in throughput by incorporating {\bf \emph{MCS Adjustment}} via re-transmissions (i.e., denoted by {\bf \emph{Baseline-Re-Tx}}) is significant, particularly, for $\mathcal{GM}$ and $\mathcal{HM}$ throughput. 

The benefits are more evident when moving from low rate NB-IoT devices to high rate 5G-NR users. The performance of sum rate (or $\mathcal{AM}$) throughput for {\bf \emph{Baseline-Rx-Tx}} is comparable with the benchmark and the DRL algorithms, but at the expense of a wasted frame causing an additional delay. 

\begin{figure*}[t!]
	\centering
	\begin{tabular}{ccc}
		\includegraphics[width=6cm,height=4cm]{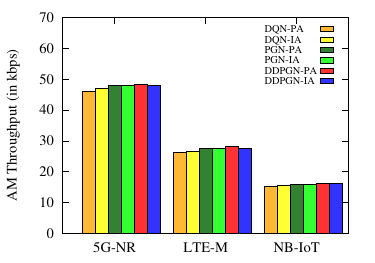}  & \hspace{-0.5cm}
		\includegraphics[width=6cm,height=4cm]{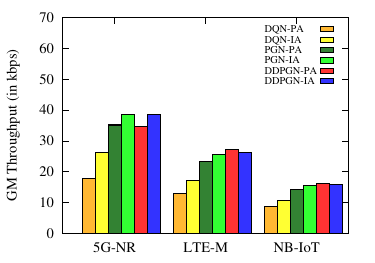} &  \hspace{-0.5cm}
		\includegraphics[width=6cm,height=4cm]{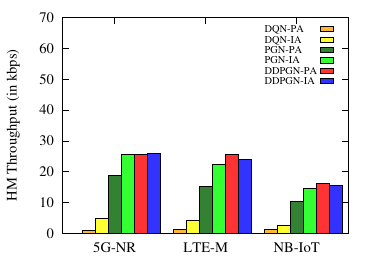} \\ 
		(a) $\Omega_{Train}$: $\mathcal{AM}$ \textit{Throughput}   &  (b) $\Omega_{Train}$: $\mathcal{GM}$ \textit{Throughput}  & (c) $\Omega_{Train}$: $\mathcal{HM}$ \textit{Throughput} \\
		\includegraphics[width=6cm,height=4cm]{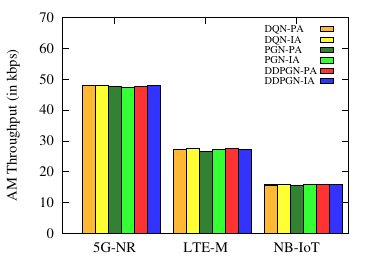}  & \hspace{-0.5cm}
		\includegraphics[width=6cm,height=4cm]{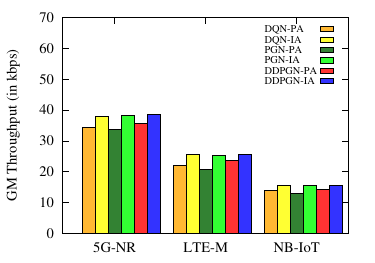}  & \hspace{-0.5cm}
		\includegraphics[width=6cm,height=4cm]{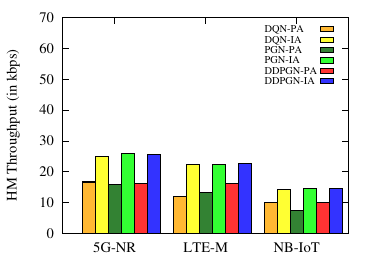} \\	
		(d) $\Omega_{Test}$: $\mathcal{AM}$ \textit{Throughput}  & (e) $\Omega_{Test}$: $\mathcal{GM}$ \textit{Throughput} & (f) $\Omega_{Test}$: $\mathcal{HM}$ \textit{Throughput}\\		
	\end{tabular}
	\centering
	\caption{Centralized Training and Testing with small scale fading}
	\label{Fading}
	
\end{figure*}

\section{DRL Training and Testing} 
\label{sec:results}
DRL algorithms are inherently designed for online learning, where an agent never stops training, and there is no separate testing phase. However, for centralized training based DRL, a training phase followed by a testing phase has been designed to see how the learned policies will be executed by the agents in a distributed setting, i.e., when centralized training has been disabled. To implement these phases, the set of realizations has been divided into two groups ($\Omega=\Omega_{Train} \cup \Omega_{Test}$) for testing and training, respectively, where $|\Omega_{Train}|=|\Omega_{Test}|= 5000$. The simulation results plotted in Fig.~\ref{No-fading} and Fig.~\ref{Fading} represent DQN testing phase after the DQN has been trained (i.e., $Q(\varsigma,\alpha;\boldsymbol{w^{q*}})$), PGN testing phase with the learned PGN (i.e., $\pi(\alpha|\varsigma; \boldsymbol{w^{\pi*}})$), and DDPGN testing phase with the learned actor (i.e., $\pi_A(\varsigma^\alpha; \boldsymbol{w^{a*}})$) and critic (i.e., $Q_C(\varsigma^c,\alpha; \boldsymbol{w^{c*}})$). Note that for centralized training, the rewards are computed via inter-agent communication using (10). 
\vspace{-0.1cm}
\subsection{Performance without Small Scale Fading}
Fig.~\ref{No-fading} compares the throughput performance of interference allocation based centralized DRL algorithms ({\bf \emph{DQN-IA, PGN-IA, DDPGN-IA}}) with the corresponding power allocation based centralized DRL algorithms ({\bf \emph{DQN-PA, PGN-PA, DDPGN-PA}}). The performance of DQN, in terms of fairness and delay (i.e., $\mathcal{GM}$ and $\mathcal{HM}$ throughput, respectively), is observed to be worst among all DRL algorithms; mainly, due to its inability to deal with outliers during the training phase. The degradation in performance can also be attributed to the amount of data required before training the DQN to find optimal parameters ($\boldsymbol{w^{q*}}$ over a set of randomly chosen sample of experiences) in Algorithm~\ref{DQN-Algo}. Once trained, the performance of {\bf \emph{DQN-PA}} is comparable to other power allocation based DRL algorithms, which can be further improved (during both training and testing phase) by adopting interference allocation in {\bf \emph{DQN-IA}}. 

The overall performance of PGN during the testing phase is comparable to the DDPGN, however, during the training phase there is a small degradation in $\mathcal{GM}$ and $\mathcal{HM}$ throughput for {\bf \emph{PGN-PA}}, which can be improved by adopting {\bf \emph{PGN-IA}}. Since, DDPGN deals with continuous action space, its performance is not affected by the discretization process; regardless of the scheme being adopted, i.e., either {\bf \emph{DDPGN-PA}} or {\bf \emph{DDPGN-IA}}), the performance is observed to be significantly better than other DRL algorithms during the training phase. However, the overall performance during the testing phase suggests that, once adequately trained, {\bf \emph{DDPGN-IA}} can outperform {\bf \emph{DDPGN-PA}} by selecting actions on the basis of interference levels as compared to the power levels.
\vspace{-0.2cm}
\subsection{Performance with Small Scale Fading}
Fig.~\ref{Fading} evaluates the performance of centralized DRL algorithms under varying channel conditions per timeslot, which are mimicked by adding small scale fading parameters. Although the sum rate (or $\mathcal{AM}$) throughput remains similar for all DRL algorithms during both training and testing phase, the additional randomness in each realization results in PGN performing slightly worse than DDPGN during the training phase. Whereas, DQN's performance during the training phase remains similar to the one without small scale fading. Here, DDPGN outperforms both PGN and DQN, mainly because it does not require discretization and thus can perform better within a large state and action space that is inherent with the induction of small scale fading. Note that varying power levels are required in each timeslot to mitigate small scale fading effects. The critic within the DDPGN can determine the Q-values using both state and action as input, where the action is generated by the actor, thereby maximizing the Q-value in each state. Whereas, the DQN learn Q-values, for all actions using only state as an input, for defining an $\epsilon$-greedy policy. 

Once the policy-based centralized DRL algorithms have been adequately trained, there is no significant difference in their performance during the testing phase. However, the overall performance (during both training and testing phase) suggests that interference allocation based centralized DRL algorithms can significantly outperform the power allocation based centralized DRL algorithms under small scale fading.  
\begin{table*}[t]
	\caption {Average Computational Latency per Timeslot for Centralized DRL Algorithms}  
	
	\begin{center}
		\begin{tabular}{|c||c|c||c|c||c|c||c|c||c|c||c|c|}
			\hline

		 	Small Scale Fading &  \multicolumn{6}{c||}{ \textbf{No}} & \multicolumn{6}{c|}{ \textbf{Yes}}\\
		 
			\hline
				 &  \multicolumn{2}{c||}{  }  &  \multicolumn{2}{c||}{ }  &  \multicolumn{2}{c||}{ } &  \multicolumn{2}{c||}{ }  &  \multicolumn{2}{c||}{}  &  \multicolumn{2}{c|}{ }\\
			User Type &  \multicolumn{2}{c||}{ \textbf{5G-NR} }  &  \multicolumn{2}{c||}{ \textbf{LTE-M}}  &  \multicolumn{2}{c||}{\textbf{NB-IoT} } &  \multicolumn{2}{c||}{ \textbf{5G-NR} }  &  \multicolumn{2}{c||}{ \textbf{LTE-M}}  &  \multicolumn{2}{c|}{\textbf{NB-IoT} }\\
			 &  \multicolumn{2}{c||}{  }  &  \multicolumn{2}{c||}{ }  &  \multicolumn{2}{c||}{ } &  \multicolumn{2}{c||}{ }  &  \multicolumn{2}{c||}{}  &  \multicolumn{2}{c|}{ }\\
			\hline
			\hline
			DRL                     &  Train &  Test & Train & Test & Train & Test &  Train &  Test & Train & Test & Train & Test\\
			Algorithm               &  (ms)  &  (ms) & (ms)  & (ms) & (ms)  & (ms)&  (ms)  &  (ms) & (ms)  & (ms) & (ms)  & (ms)\\
			\hline		 
			\textbf{DQN-PA} & 0.88   & 0.71 & 0.97 & 0.78 & 0.89  & 0.76 & 0.86  & 0.69 & 0.88 & 0.72 & 0.88 & 0.73 \\
			\hline
			\textbf{DQN-IA} & 0.89 &  0.73 & 0.90 & 0.71 & 0.89 & 0.74 &  0.90 & 0.72  & 0.90 & 0.71 & 0.90 & 0.73\\
			\hline
			\textbf{PGN-PA} & 4.50    & 0.71  & 4.57  & 0.71 & 4.51  & 0.74 & 4.45 & 0.69 & 4.55 & 0.734 & 4.51& 0.74 \\
			\hline
			\textbf{PGN-IA} & 4.74 & 0.81  & 4.51 & 0.70 & 4.52& 0.75 & 4.50 & 0.73 & 4.73 & 0.74 & 4.68 & 0.75\\
			\hline
			\textbf{DDPGN-PA} & 3.8   & 0.71 & 3.72  & 0.68 & 3.75  & 0.75& 3.77 & 0.70 & 3.72 & 0.69& 3.75& 0.75\\
			\hline
			\textbf{DDPGN-IA} & 3.88  & 0.73  & 3.79 & 0.71 & 3.75 & 0.71 & 3.84  & 0.71 & 3.75 & 0.70 & 3.74 & 0.74\\
			\hline
			
		\end{tabular}
	\vspace{-0.3cm}
	\end{center}
	\label{timecost}
\end{table*}
\begin{table}[t]
	\caption {Average Computational Latency per Timeslot for Edge Learning based DRL Algorithms}  

	\begin{center}
		\begin{tabular}{|c||c|c||c|c||c|c|}
			\hline
			
			Small Scale Fading &  \multicolumn{6}{c|}{ \textbf{No}} \\
				
						\hline
									 &  \multicolumn{2}{c||}{ }  &  \multicolumn{2}{c||}{ }  &  \multicolumn{2}{c|}{ } \\
			User Type &  \multicolumn{2}{c||}{ \textbf{5G-NR} }  &  \multicolumn{2}{c||}{ \textbf{LTE-M}}  &  \multicolumn{2}{c|}{\textbf{NB-IoT} } \\
				 &  \multicolumn{2}{c||}{ }  &  \multicolumn{2}{c||}{ }  &  \multicolumn{2}{c|}{ } \\
			\hline
			\hline
			DRL  &   \multicolumn{2}{c||}{Train} &  \multicolumn{2}{c||}{Train} &   \multicolumn{2}{c|}{Train}\\
			Algorithm   &   \multicolumn{2}{c||}{(ms)}  &  \multicolumn{2}{c||}{(ms)}  &  \multicolumn{2}{c|}{(ms)}\\
			\hline		 
		
			\textbf{DQN-IA} & \multicolumn{2}{c||}{ 0.85 } & \multicolumn{2}{c||}{ 0.78 } & \multicolumn{2}{c|}{ 0.73 }\\
			\hline
				
			\textbf{PGN-IA} & \multicolumn{2}{c||}{2.07 } & \multicolumn{2}{c||}{ 1.96 } & \multicolumn{2}{c|}{1.92 } \\
			\hline
		
			\textbf{DDPGN-IA} & \multicolumn{2}{c||}{ 3.26} & \multicolumn{2}{c||}{3.07 } & \multicolumn{2}{c|}{ 3.02 } \\
			
			\hline
			
		\end{tabular}
	
	\end{center}
		\vspace{-0.4cm}
	\label{edge-timecost}
\end{table}
\begin{figure}[t!]
	\centering
	\begin{tabular}{cc}
		\includegraphics[width=9cm,height=4.8cm]{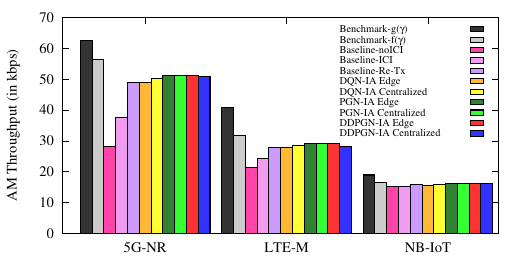}  \\ 
		(a) $\Omega_{Train}$: $\mathcal{AM}$ \textit{Throughput}     \\
		\includegraphics[width=9cm,height=4.8cm]{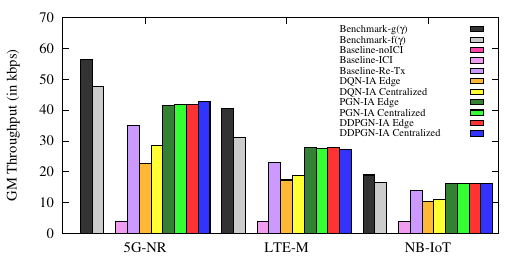}   \\	
		(b) $\Omega_{Train}$: $\mathcal{GM}$ \textit{Throughput} \\
		\includegraphics[width=9cm,height=4.8cm]{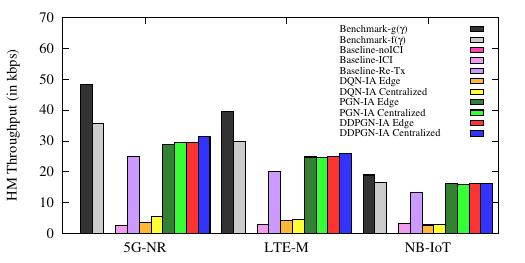}  \\ 
		(c) $\Omega_{Train}$: $\mathcal{HM}$ \textit{Throughput}    		
	\end{tabular}
	\centering
	\caption{Edge Learning without small scale fading}
	\label{Edge}

\end{figure}
\newpage
\subsection{Comparison with Edge Learning}
For edge learning, the set of realizations for training (i.e., $|\Omega_{Train}|= 5000$) has been used only. The simulation results plotted in Fig.~\ref{Edge} represent the edge learning phase when the rewards for each agent are computed using (9). The comparison with the training phase of centralized DRL algorithms show no significant change in average throughput for policy-based DRL algorithms. However, there is a degradation in $\mathcal{GM}$ as well as $\mathcal{HM}$ throughput for value-based DQN edge learning algorithms.

\subsection{Computational Latency Evaluation}
\label{sec:latency}
Low computational delay is a desirable feature for real-time systems, therefore, the computational latency of centralized and edge learning based DRL algorithms have been compared in Table~\ref{timecost} and Table~\ref{edge-timecost}, respectively. 

For both power allocation and interference allocation based DRL algorithms, it can be seen that DQN is more time efficient than PGN and DDPGN, since, it is meant for approximating a function rather then determining a policy. The interference allocation based DRL algorithms (i.e., {\bf \emph{DQN-IA, PGN-IA, DDPGN-IA}}) incur slightly higher computational cost when compared to the corresponding power allocation based DRL algorithms (i.e., {\bf \emph{DQN-PA, PGN-PA, DDPGN-PA}}), i.e.,  due to the additional computations that are required for allocating interference before allocating power. 

The centralized training phase for both PGN and DDPGN requires more computational time, mainly because they require higher number of parameters as compared to the DQN. Once trained, the centralized DRL algorithms have almost similar computational latency during the testing phase. In contrast, the computational latency for edge learning based DRL algorithms in Table~\ref{edge-timecost} is much lower than their centralized counterparts.